\documentclass[epj]{webofc}
\usepackage[utf8]{inputenc}
\usepackage[varg]{txfonts}   
\usepackage{booktabs}
\usepackage{xcolor}
\definecolor{darkred}{rgb}{0.4,0.0,0.0}
\definecolor{darkgreen}{rgb}{0.0,0.4,0.0}
\definecolor{darkblue}{rgb}{0.0,0.0,0.4}
\usepackage[bookmarks,linktocpage,colorlinks,
    linkcolor = darkred,
    urlcolor  = darkblue,
    citecolor = darkgreen]{hyperref}
\usepackage{subfigure}
\usepackage{physics}
\usepackage{dsfont}
\usepackage[percent]{overpic}
\usepackage{rotating}
\wocname{EPJ Web of Conferences}
\woctitle{Lattice2017}
\begin{document}
%
\selectlanguage{english}
\title{%
Computation of hybrid static potentials in SU(3) lattice gauge theory
}
\author{%
\firstname{Christian} \lastname{Reisinger}\inst{1}\fnsep\thanks{Speaker, \email{reisinger@th.physik.uni-frankfurt.de}} \and
\firstname{Stefano} \lastname{Capitani}\inst{1} \and
\firstname{Owe} \lastname{Philipsen}\inst{1} \and
\firstname{Marc} \lastname{Wagner}\inst{1}
}
\institute{%
Institut f\"ur Theoretische Physik, Goethe-Universit\"at Frankfurt am Main, Max-von-Laue-Stra{\ss}e 1, D-60438 Frankfurt am Main, Germany
}
\abstract{%
We compute hybrid static potentials in SU(3) lattice gauge theory. We present a method to generate a large set of suitable creation operators with defined quantum numbers from elementary building blocks. We show preliminary results for several channels and discuss, which structures of the gluonic flux tube seem to be realized by the ground states in these channels.
}
\maketitle
\section{Introduction}\label{intro}

The existence of states containing gluonic excitations is suggested by QCD. These excitations
contribute in a non-trivial way to the properties of bound states. For example mesonic
states with gluonic excitations, called hybrid mesons, can carry quantum numbers different from
those in the quark model. A better understanding of exotic matter like hybrid mesons is
important to further improve our understanding of the strong interactions. The search for exotic matter
is also a popular topic in current experiments and a theoretical investigation is essential to analyze experimental data.

In this work we discuss, how to obtain hybrid static potentials relevant for hybrid mesons with heavy quarks using lattice computations in SU(3) gauge theory. Focus is put on finding a suitable set of creation operators,
to obtain trial states with large overlaps to the corresponding hybrid static potential ground states. We show, how
the trial states are generated and outline our procedure to find suitable operators. Finally, we show first results for hybrid static potentials with absolute angular momenta $L = 0,1,2$ with respect to the axis of separation of the static quark antiquark pair and compare to results from the literature \cite{Juge:1997nc}. For further existing lattice studies cf.\ \cite{Peardon:1997jr,Juge:1997ir,
Morningstar:1998xh,Michael:1998tr,Michael:1999ge,Toussaint:1999kh,
Bali:2000vr,Morningstar:2001nu,Juge:2002br,
Michael:2003ai,Juge:2003qd,Michael:2003xg,Bali:2003jq,Wolf:2014tta}, for a recent effective field theory description cf.\ \cite{Berwein:2015vca}.

\section{Hybrid mesons on the lattice}\label{sec:1}

A hybrid static potential is a static potential of a quark antiquark pair with additional
gluonic contributions to its quantum numbers. To obtain static
potentials from lattice computations in SU(3) gauge theory, we generate an ensemble of gauge configurations using the Wilson gauge action and compute Wilson loop-like correlation functions. Hybrid static potentials are then obtained from the corresponding effective masses.
To implement gluonic excitations of hybrid mesons in our trial states, we replace the spatial Wilson lines of
Wilson loops by shapes more complicated than a straight
line.

The quantum numbers of hybrid static potentials are the following (for a more detailed discussion cf.\ e.g.\ \cite{Bali:2005fu,Bicudo:2015kna}).
\begin{itemize}
\item Absolute angular momentum with respect to the axis of separation of the static quark antiquark pair $L = 0,1,2,\ldots$.

\item $Q_{P C} = +,-$ corresponding to the operator $\mathcal{P} \circ \mathcal{C}$, i.e.\ the combination of parity and charge conjugation.

\item $P_x = +,-$ corresponding to the operator $\mathcal{P}_x$, which corresponds to the spatial reflection along an axis perpendicular to the axis of separation of the static quark antiquark pair.
\end{itemize}
It is conventional to write $L = \Sigma,\Pi,\Delta$ instead of $L = 0,1,2$ and $Q_{P C} = g,u$ instead of $Q_{P C} = +,-$. Note that for angular momentum $L > 0$ the spectrum is degenerate with respect to $P_x = +$ and $P_x = -$. The labeling of states is thus $L^{P_x}_{Q_{PC}}$ for $L = 0 = \Sigma$ and $L_{Q_{PC}}$ for $L > 0$.


\subsection{Angular momentum $L$}\label{ssec:1-2}

We place the quark and the antiquark at positions $\mathbf{r}_q=(0,0,+r/2)$ and $\mathbf{r}_{\bar{q}}=(0,0,-r/2)$, i.e.\ separate them along the $z$ axis. In the following we only write the $z$ coordinate explicitly.

In a first step we consider trial states, which read in the continuum
\begin{equation}
\ket{\Psi_\text{Hybrid}}_L = \int_0^{2\pi} d\varphi \, \operatorname{exp}(iL\varphi) \hat{R}
(\varphi) \hat{O} \ket{\Omega},
\end{equation}
where $\ket{\Omega}$ is the vacuum and $\hat{R}(\varphi)$ denotes a rotation around angle $\varphi$ with respect to the $z$ axis.
\begin{equation}
\label{eq:psiS}
\hat{O} \ket{\Omega} = \bar{q}(-r/2) S(-r/2,+r/2) q(+r/2) \ket{\Omega},
\end{equation}
where $S(-r/2,+r/2)$ connects the quark and the antiquark in a gauge invariant way and has a non-trivial shape and, thus, generates gluonic excitations.
Such trial states have defined angular momentum $L$.

The corresponding lattice expression is
\begin{equation}
\label{eq:L}
\ket{\Psi_\text{Hybrid}}_L = \sum_{k = 0}^3 \operatorname{exp}\bigg(iLk\frac{\pi}{2}\bigg)
\hat{R}\bigg(k\frac{\pi}{2}\bigg)\hat{O}\ket{\Omega},
\end{equation}
where the rotation angles are restricted to multiples of $\pi/2$ and $S(-r/2,+r/2)$ is a function of the link variables. E.g.\ for $L=0$
\begin{equation}
\ket{\Psi_\text{Hybrid}}_{L=0} = \bigg[1 + \hat{R}\bigg(\frac{\pi}{2}\bigg) + \hat{R}(\pi)
+ \hat{R}\bigg(\frac{3\pi}{2}\bigg)\bigg] \hat{O} \ket{\Omega} ,
\end{equation}
i.e.\ we have to compute
Wilson loops, where the spatial Wilson lines are a sum over rotations of the shape $\hat{O}$ with weight factors according to eq.\ (\ref{eq:L}). Note that, due to the restriction to cubic rotations, the lattice trial states do not have defined angular momentum, but contain also higher angular momentum excitations.


\subsection{$Q_{P C}$ and $P_x$}

It is straightforward to show
\begin{equation}
\mathcal{P} \circ \mathcal{C} \hat{O} \ket{\Omega} = \mathcal{P} \circ \mathcal{C} \bar{q}(-r/2) S(-r/2,+r/2) q(+r/2) \ket{\Omega} = \bar{q}(-r/2) [S_\mathcal{P}](-r/2,+r/2) q(+r/2) \ket{\Omega} ,
\end{equation}
where $S_\mathcal{P}$ is the spatial reflection of $S$ with respect to the midpoint of the separation axis. Consequently, one has to include both $S$ and $S_\mathcal{P}$ in the final operator, to obtain a trial state with defined $Q_{P C}$. Similarly,
\begin{equation}
\mathcal{P}_x \hat{O} \ket{\Omega} = \mathcal{P}_x \bar{q}(-r/2) S(-r/2,+r/2) q(+r/2) \ket{\Omega} = \bar{q}(-r/2) [S_{\mathcal{P}_x}](-r/2,+r/2) q(+r/2) \ket{\Omega} ,
\end{equation}
where $S_{\mathcal{P}_x}$ is the spatial reflection of $S$ along an axis perpendicular to the axis of separation as defined above.

\begin{figure}[thb]
\centering
\includegraphics[width=7cm,clip]{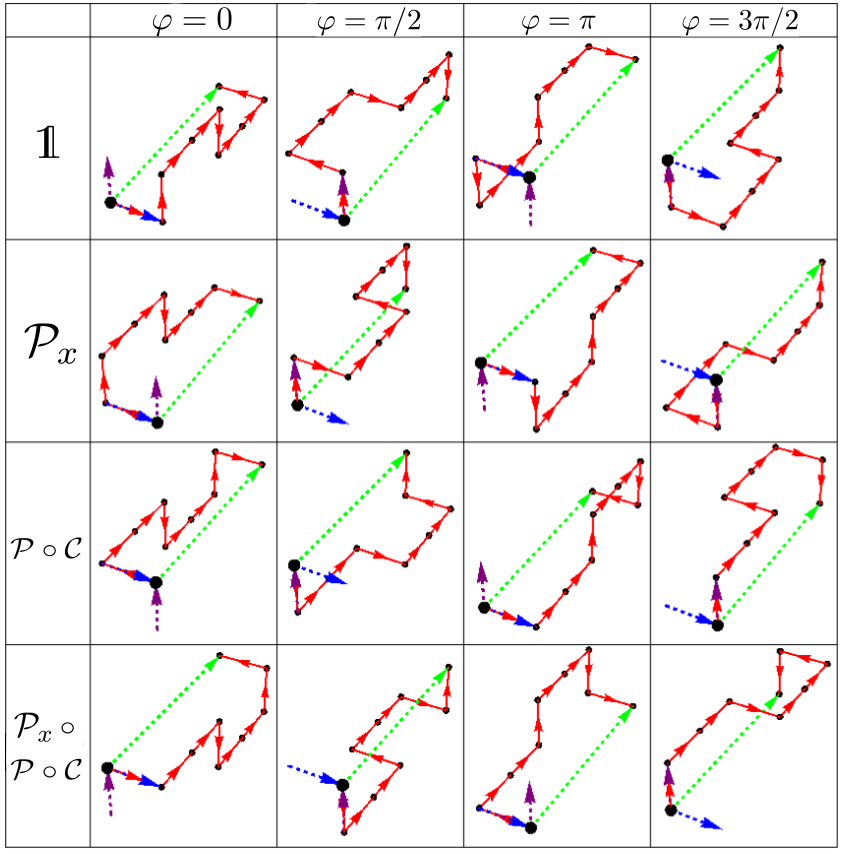}
\caption{\label{fig:trafotable}The terms appearing in the construction of trial state via eq.\ (\ref{eq:trialstate}) for an exemplary shape $S(-r/2,+r/2)$ of the spatial Wilson line (top left). The columns correspond to rotations and the rows to applications of the operators $\mathcal{P}_x$ and $\mathcal{P} \circ \mathcal{C}$. Continuous red lines represent link variables, dotted lines the $z$ axis and black dots the lattice sites.}
\end{figure}

To construct a trial state, which has defined quantum numbers $L$, $Q_{P C}$ and $P_x$, we start with a state
with defined angular momentum $L$, eq.\ (\ref{eq:L}), and project that state onto the subspace of eigenstates of the operators $\mathcal{P} \circ \mathcal{C}$ and $\mathcal{P}_x$ characterized by $Q_{P C}$ and $P_x$:
\begin{eqnarray}
\nonumber & & \hspace{-0.7cm} \ket{\Psi_\text{Hybrid}}_{L,Q_{P C},P_x} = \mathds{P}_{P_x}\, \mathds{P}_{PC}\, \ket{\Psi_\text{Hybrid}}_L = \\
\nonumber & & = \Big(1 + P_x \mathcal{P}_x + Q_{PC}\mathcal{P}\circ\mathcal{C} + P_x Q_{PC} \mathcal{P}_x
\mathcal{P}\circ\mathcal{C}\Big) \sum_{k=0}^3 \operatorname{exp}\bigg(iLk\frac{\pi}{2}\bigg) \hat{R}\bigg(k\frac{\pi}{2}\bigg) \hat{O} \ket{\Omega} \equiv \\
\label{eq:trialstate} & & \equiv \bar{q}(-r/2) a_S[L_{Q_{P C}}^{P_x}](-r/2,+r/2) q(+r/2) \ket{\Omega}
\end{eqnarray}
with projectors
\begin{equation}
\mathds{P}_{P C} = \frac{1}{2}(1 + Q_{PC} \mathcal{P} \circ \mathcal{C}) \quad , \quad \mathds{P}_{P_x} = \frac{1}{2} (1 + P_x \mathcal{P}_x) .
\end{equation}
Note that for a shape $S(-r/2,+r/2)$, which cannot be used to construct a trial state with a specific
choice of quantum numbers $(L,Q_{PC},P_x)$, the trial state (\ref{eq:trialstate})
automatically vanishes, i.e.\ $\ket{\Psi_\text{Hybrid}} = 0$. In practice, we use eq.\ (\ref{eq:trialstate}) to quickly generate creation operators with defined $(L,Q_{PC},P_x)$ from any given input shape $S$ (cf.\ Figure~\ref{fig:trafotable} for a graphical illustration of an example).

\section{Numerical results}

To obtain hybrid static potentials, we have computed Wilson loop-like correlation functions
using the shapes $S$ according to eq.\ (\ref{eq:trialstate}) as spatial Wilson lines. These computations have been performed on gauge link configurations generated with the standard Wilson
gauge action and the Chroma QCD library \cite{Edwards:2004sx}. We have used lattices of size $24^3 \times 48$ and gauge coupling $\beta = 6.0$, which corresponds to a
lattice spacing $a \approx 0.093\, \text{fm}$, when identifying $r_0$ with $0.5 \, \textrm{fm}$ \cite{Koma:2006si}.


Hybrid static potentials can be obtained with smaller statistical errors, when using trial states with larger overlaps to the corresponding energy eigenstates of interest, since then effective masses exhibit plateaus at smaller temporal separations. To construct such trial states, we have considered many different shapes $S(-r/2,+r/2)$ and studied the overlaps of the corresponding operators to the energy eigenstates of interest. In other words, we have investigated, which structure of the gluonic flux tube between the quark and the antiquark is realized for each hybrid potential.

APE smearing and HYP smearing have also been used to improve the signal quality (cf.\ e.g.\ \cite{Jansen:2008si} for detailed equations).


\subsection{Optimization of APE smearing}

\begin{figure}[thb]
\centering
\hspace*{1.3cm}
\begin{overpic}[width=5cm]{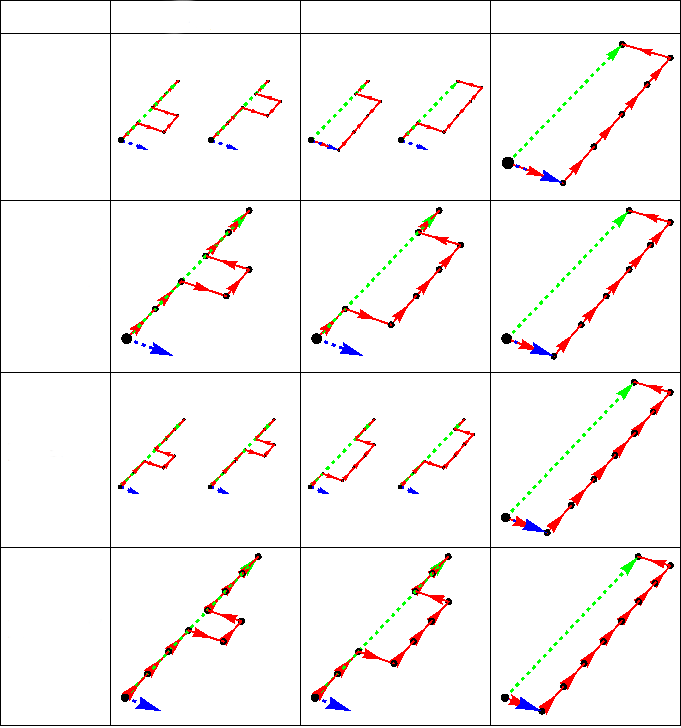}
\put (25,96.5) {\scriptsize{$S_{0,1}$}}
\put (51,96.5) {\scriptsize{$S_{0,2}$}}
\put (77,96.5) {\scriptsize{$S_{0,3}$}}
\put (0.5,82) {\scriptsize{$r/a = 4$}}
\put (0.5,60) {\scriptsize{$r/a = 5$}}
\put (0.5,36) {\scriptsize{$r/a = 6$}}
\put (0.5,11) {\scriptsize{$r/a = 7$}}
\end{overpic}%
\hspace*{0.2cm}
\begin{overpic}[width=8cm]{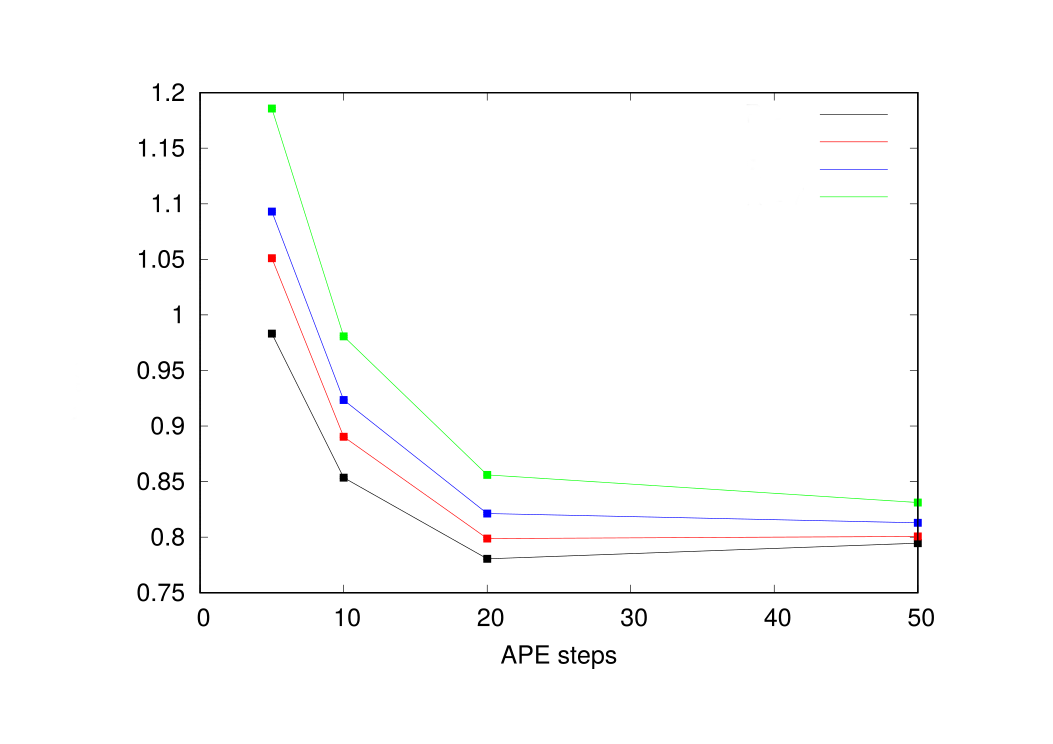}
\put (8,30) {\begin{sideways}\scriptsize{$V_\text{eff}(t/a=1)\,a$}\end{sideways}}
\put (70,59) {\tiny{$r/a = 4$}}
\put (70,56.4) {\tiny{$r/a = 5$}}
\put (70,53.8) {\tiny{$r/a = 6$}}
\put (70,51.2) {\tiny{$r/a = 7$}}
\put (50,65) {$S_{0,1}$}
\end{overpic}
\hspace*{-0.5cm}
\begin{overpic}[width=8cm]{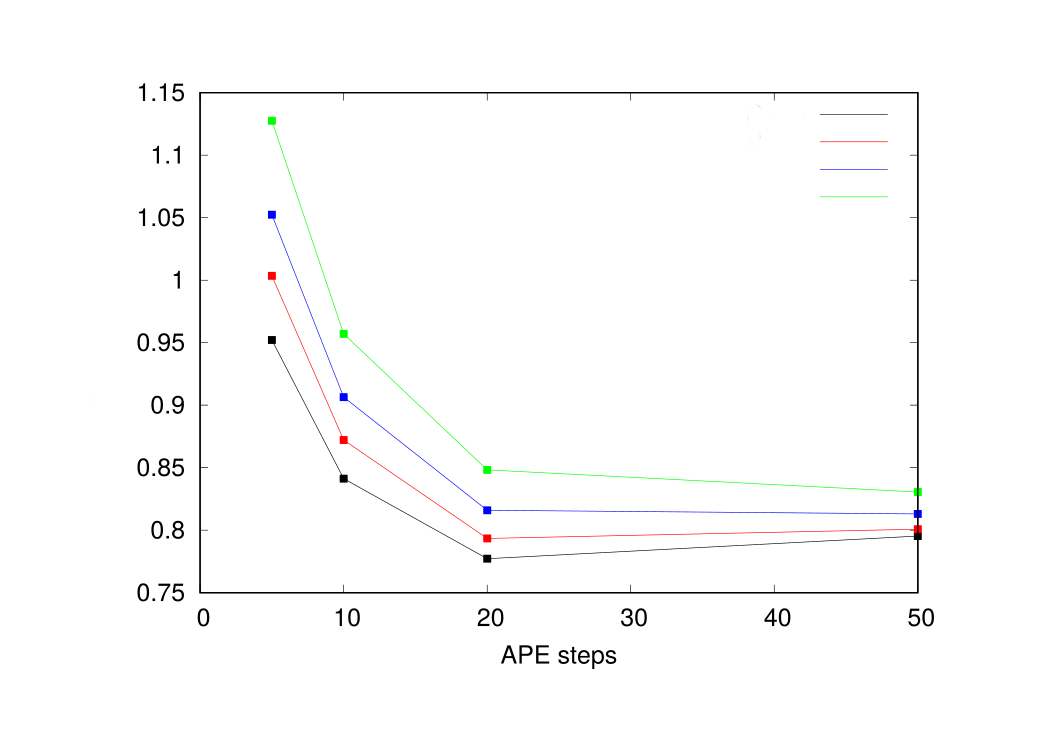}
\put (8,30) {\begin{sideways}\scriptsize{$V_\text{eff}(t/a=1)\,a$}\end{sideways}}
\put (70,59) {\tiny{$r/a = 4$}}
\put (70,56.4) {\tiny{$r/a = 5$}}
\put (70,53.8) {\tiny{$r/a = 6$}}
\put (70,51.2) {\tiny{$r/a = 7$}}
\put (50,65) {$S_{0,2}$}
\end{overpic}%
\hspace*{-1cm}
\begin{overpic}[width=8cm]{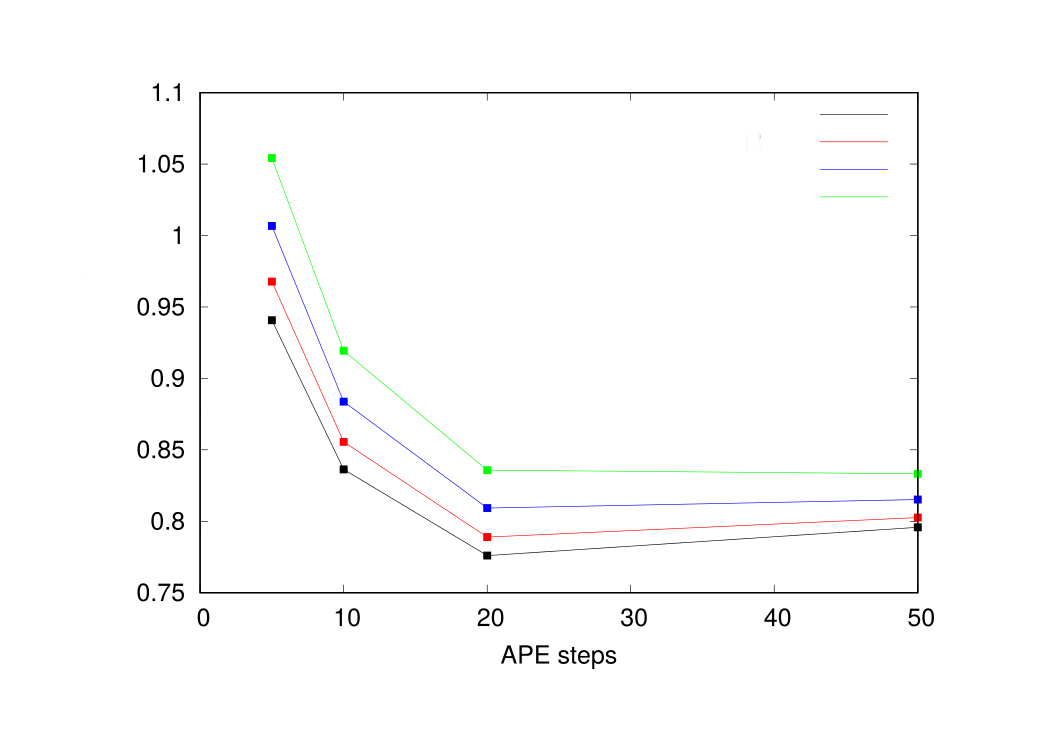}
\put (8,30) {\begin{sideways}\scriptsize{$V_\text{eff}(t/a=1)\,a$}\end{sideways}}
\put (70,59) {\tiny{$r/a = 4$}}
\put (70,56.4) {\tiny{$r/a = 5$}}
\put (70,53.8) {\tiny{$r/a = 6$}}
\put (70,51.2) {\tiny{$r/a = 7$}}
\put (50,65) {$S_{0,3}$}
\end{overpic}
\caption{Operators used for the optimization of the number of APE smearing steps (two shapes
in a single cell imply the average ``$(\text{left shape} + \text{right shape}) / 2$'') and
associated effective masses at $t = a$ and $N_\text{APE}=5,10,20,50$ (top right: $S_{0,1}$;
bottom left: $S_{0,2}$; bottom right: $S_{0,3}$).}
\label{fig:APEops}
\end{figure}
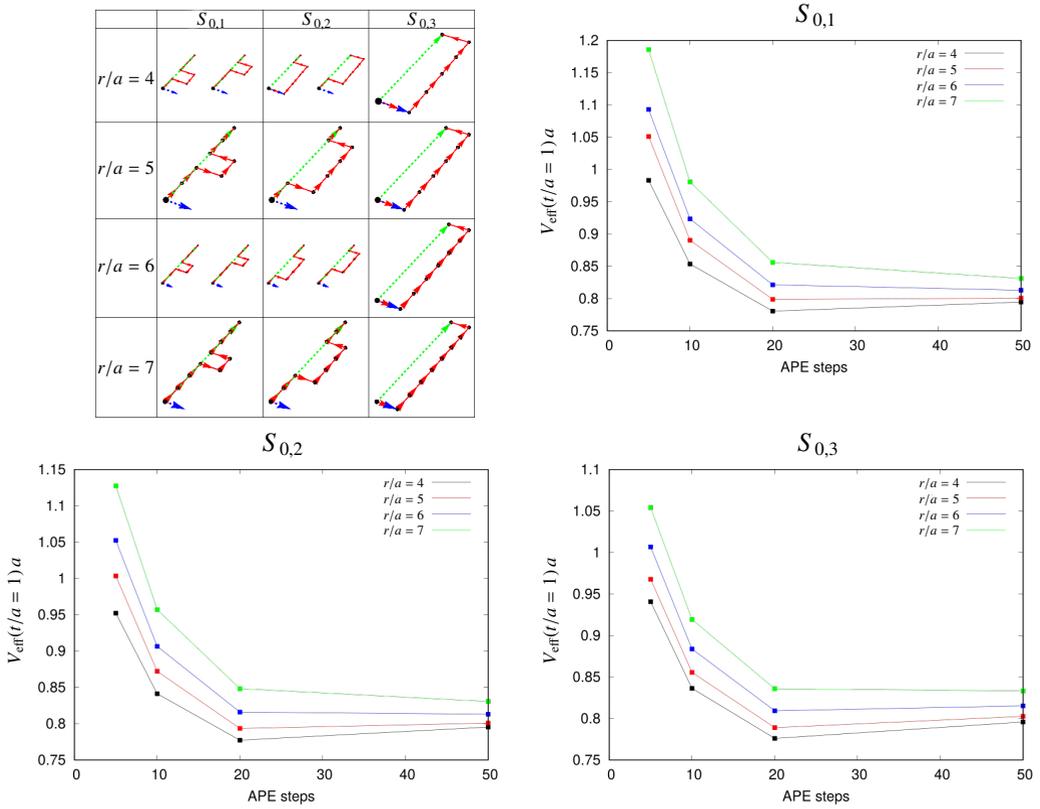

In an initial step we have considered a simple \emph{staple} shape (cf.\ Figure~\ref{fig:APEops}) with varying
extension along the quark antiquark separation axis and have computed effective masses at small temporal separation $t = a$ for different
quark antiquark separations and different numbers of APE-smearing steps $N_\text{APE}$.
The optimal choice for $N_\text{APE}$ corresponds to the lowest value of the effective mass at $t = a$, as it implies that a plateau is reached at earlier times, where the signal-to-noise ratio is still large.

We observe a minimum at $N_\text{APE} = 20$ for $r / a = 4, 5$ for all operators
and only a small change of the effective mass at $N_{APE} = 50$. We also see that the effective mass
for the longest shape $S_{0,3}$ is slightly lower than for the shorter staples and, thus, results in a larger overlap of the corresponding trial state with the ground state. Since we are particularly interested in the region of small quark antiquark separations, we decided to use $N_\text{APE} = 20$ for all computations presented in the following.


\subsection{\label{SEC_3_2}Optimization of operator shapes}

\begin{figure}[thb]
\centering
\begin{overpic}[scale=0.23]{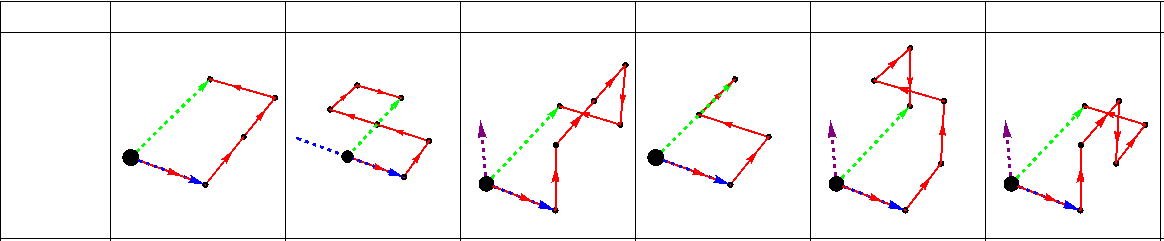}
\put (0.5,8) {\scriptsize{$r/a = 2$}}
\put (16,18.5) {\scriptsize{$S_1$}}
\put (31,18.5) {\scriptsize{$S_2$}}
\put (46,18.5) {\scriptsize{$S_3$}}
\put (61,18.5) {\scriptsize{$S_4$}}
\put (76,18.5) {\scriptsize{$S_5$}}
\put (92,18.5) {\scriptsize{$S_6$}}
\end{overpic}\\
\vspace*{0.5cm}
\begin{overpic}[scale=0.23]{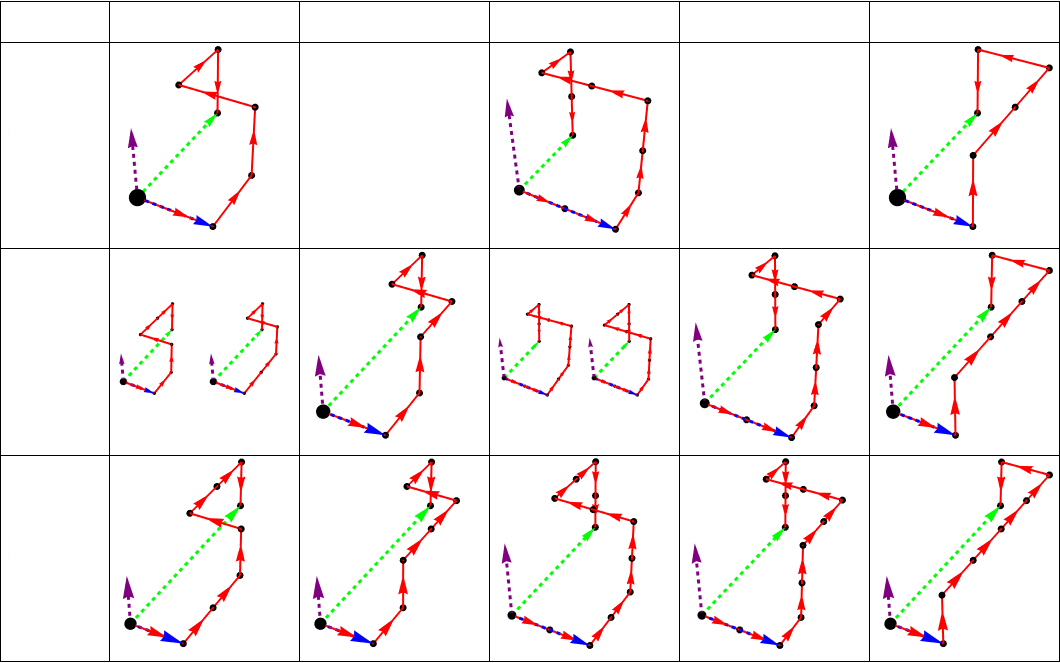}
\put (0.5,48) {\scriptsize{$r/a = 2$}}
\put (0.5,29) {\scriptsize{$r/a = 3$}}
\put (0.5,9) {\scriptsize{$r/a = 4$}}
\put (17,59.5) {\small{$S_{5,0}$}}
\put (35,59.5) {\small{$S_{5,1}$}}
\put (53,59.5) {\small{$S_{5,2}$}}
\put (71,59.5) {\small{$S_{5,3}$}}
\put (89,59.5) {\small{$S_{5,4}$}}
\end{overpic}
\caption{\label{FIG003}Basic operator shapes $S_1$ to $S_6$ (top) and an example for variations of the extensions of shape $S_5$ (bottom).}
\end{figure}

Then we have computed effective masses for a large set of operators using a variety of
different shapes $S(-r/2,+r/2)$. We have considered six significantly different basic shapes (cf.\ Figure~\ref{FIG003}, top), but have also investigated variations by slightly varying their extensions:
\begin{itemize}
\item Extensions along the axis of separation, i.e.\ the number of consecutive links in
$z$ direction (a shape can have multiple $z$ extensions corresponding to sub-shapes separated by links in
$x$ or $y$ direction).

\item Extensions along the $x$ or $y$ axes (again a shape can have multiple $x$ or $y$ extensions corresponding to different sub-shapes).
\end{itemize}
An example for such a variations of the extensions of shape $S_5$ is shown in Figure~\ref{FIG003}, bottom.

\begin{figure}[thb]
\centering
\begin{overpic}[width=\textwidth]{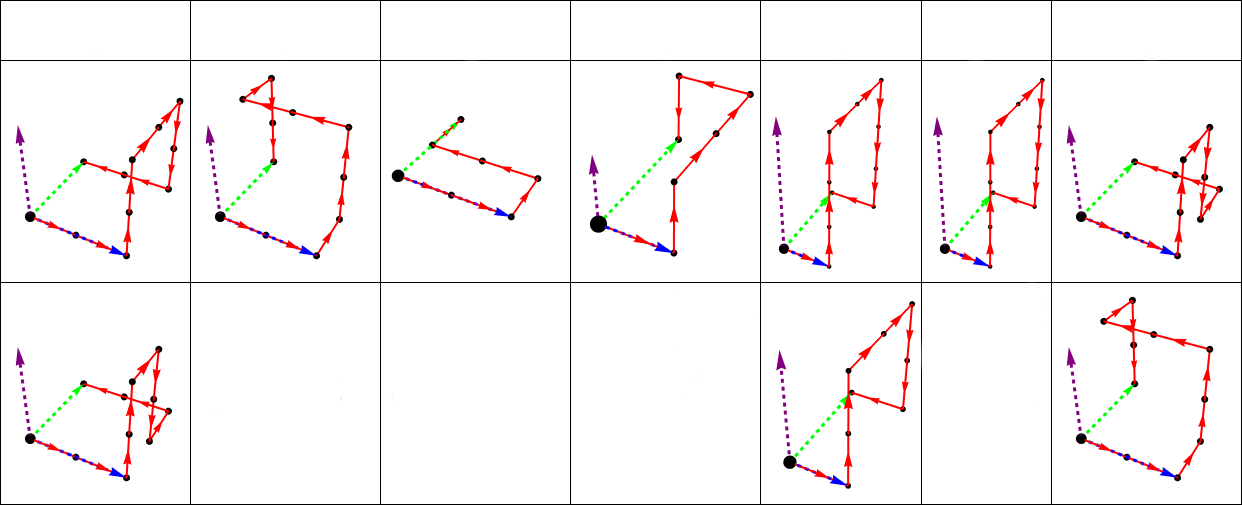}
\put (7,37.5) {$\Sigma^-_g$}
\put (22,37.5) {$\Sigma^-_u$}
\put (37,37.5) {$\Sigma^+_u$}
\put (53,37.5) {$\Pi_g$}
\put (67,37.5) {$\Pi_u$}
\put (79,37.5) {$\Delta_g$}
\put (92,37.5) {$\Delta_u$}
\end{overpic}
\caption{\label{fig:best}Operator shapes, which generate large ground state overlaps at quark antiquark separation $r/a = 2$. In some cases two shapes are shown, since the corresponding effective masses at $t = a$ are compatible within statistical errors.}
\end{figure}
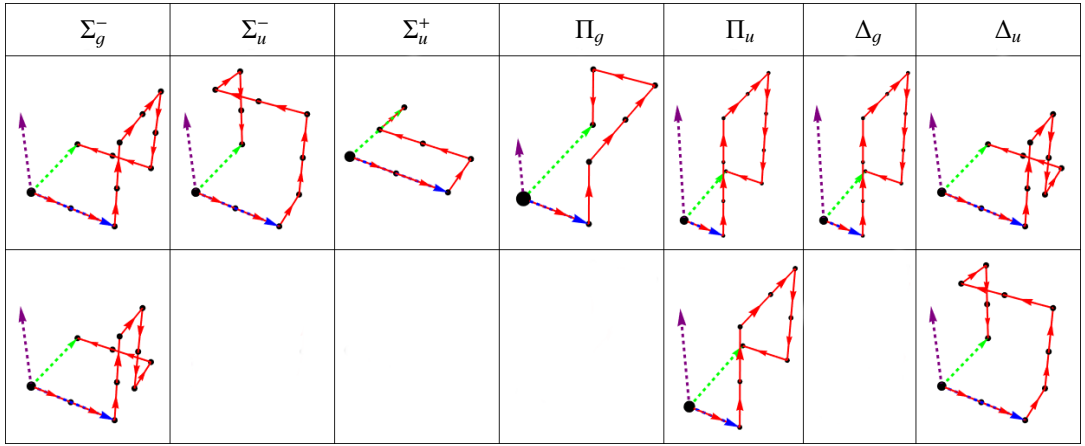

We observe that increasing the extensions of a shape typically results in a significantly larger ground state overlap. However, the optimal values for the extensions are different for different shapes and quantum numbers and are also weakly dependent on the spatial quark separation. Exceptions are $S_5$ for the states $\Delta_g, \Delta_u$ and $S_6$ for the state $\Sigma^-_u$, where more local shapes have led to larger ground state overlaps and, hence, to better results. Fig.~\ref{fig:best} shows for seven different low-lying hybrid potentials the corresponding optimized shapes, for which the effective mass at $t = a$ is quite low. It is expected that the gluonic flux tubes associated with these hybrid potentials exhibit a similar geometry.

\subsection{Hybrid static potentials}

Finally we have computed hybrid static potentials on 700 gauge link configurations using the basic operator shapes
$S_j$, $j=1,\dots,6$. In this computation variations of the extensions have not been considered, as the optimization discussed in section~\ref{SEC_3_2} is still ongoing. The hybrid static potentials have been obtained by solving generalized eigenvalue problems (cf.\ e.g.\ \cite{Blossier:2009kd}) for correlation matrices
\begin{equation}
C_{j k}(t) = \operatorname{Tr}\Big\{ (a_{S_j})_{t_0}[L^{P_x}_{Q_{PC}}]
T(t_0,t_1,\mathbf{r}_q)\left((a_{S_k})_{t_1}[L^{P_x}_{Q_{PC}}]\right)^\dagger
T^\dagger(t_0, t_1, \mathbf{r}_{\bar{q}}) \Big\}
\end{equation}
with $t = t_1-t_0$ and $(a_{S_j})_t[L^{P_x}_{Q_{PC}}]$ as defined in eq.\ (\ref{eq:trialstate}), where $t$ denotes the time argument of the corresponding spatial links. $T(t_0,t_1,\mathbf{r})$ is the HYP2 smeared temporal
Wilson line from time $t_0$ to time $t_1$ at spatial position $\mathbf{r}$.

\begin{figure}[thb]
\centering
\subfigure[Our results]{
\includegraphics[width=.3\textwidth]{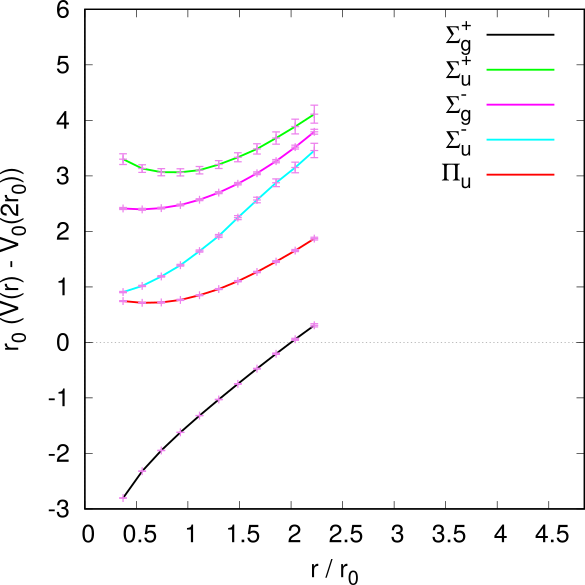}
\includegraphics[width=.3\textwidth]{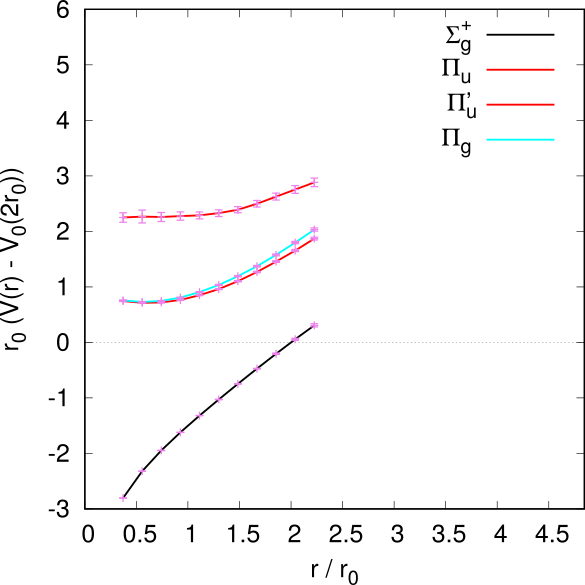}
\includegraphics[width=.3\textwidth]{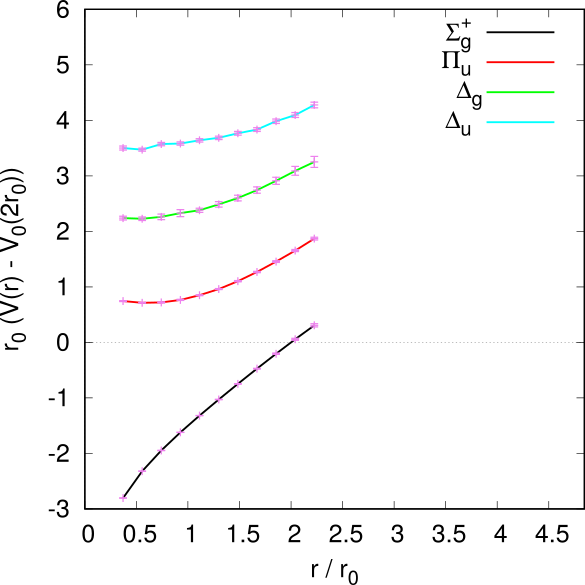}
}
\subfigure[Results taken from \cite{Juge:1997nc}]{
\includegraphics[width=.3\textwidth]{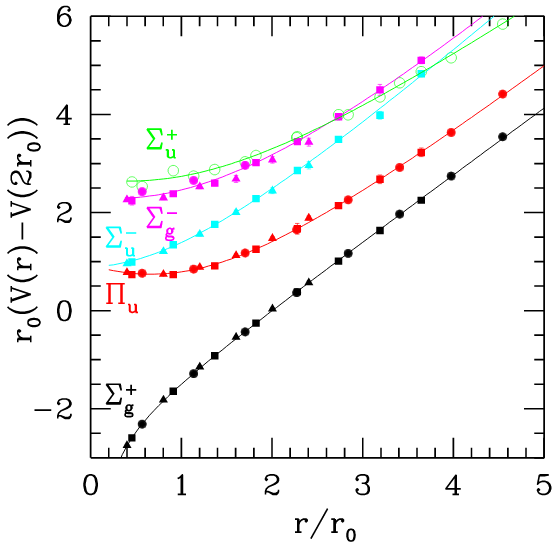}
\includegraphics[width=.3\textwidth]{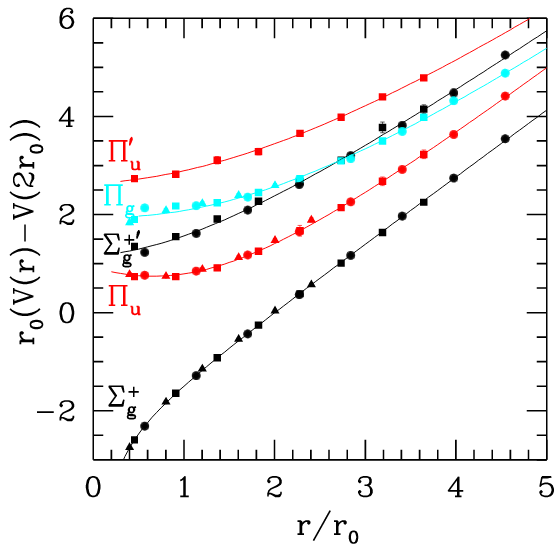}
\includegraphics[width=.3\textwidth]{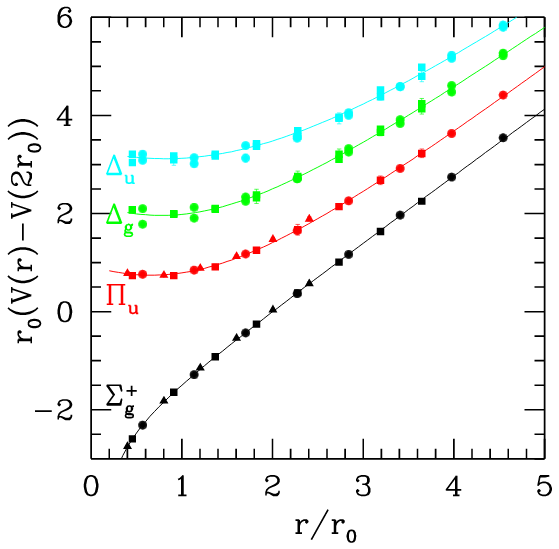}
}
\caption{Static hybrid potentials for angular momentum $L=0,1,2$ (left to right).
Our results (top) compared to the results from \cite{Juge:1997nc} (bottom).}
\label{fig:potentials}
\end{figure}

In Figure~\ref{fig:potentials} we show our current results in comparison with results from the literature \cite{Juge:1997nc}. We note that at our current level of statistical accuracy it is difficult to unambiguously identify effective mass plateaus. The shown potentials have been obtained from fits in a region of $t$, where statistical errors are still small. Keeping this in mind our results are in fair agreement with those of \cite{Juge:1997nc} with the exception of the $\Pi_g$ potential. The origin of this discrepancy is not clear and will be subject of further investigations.

\section*{Acknowledgements}

C.R.\ acknowledges support by a Karin and Carlo Giersch Scholarship of the Giersch foundation. O.P.\ and M.W.\ acknowledge support by the DFG (German Research Foundation), grants PH 158/4-1 and WA 3000/2-1. M.W.\ acknowledges support by the Emmy Noether Programme of the DFG, grant WA 3000/1-1.

This work was supported in part by the Helmholtz International Center for FAIR within the framework of the LOEWE program launched by the State of Hesse.

Calculations on the LOEWE-CSC and on the on the FUCHS-CSC high-performance computer of the Frankfurt University were conducted for this research. We would like to thank HPC-Hessen, funded by the State Ministry of Higher Education, Research and the Arts, for programming advice.


%

\end{document}